\documentclass{sf2a-conf}
\usepackage{graphicx}
\usepackage{txfonts}
\usepackage[latin1]{inputenc}
\usepackage[T1]{fontenc}  
\begin{document}
\TitreGlobal{SF2A 2007}

\title{Numerical methods comparison for protostellar collapse calculations.}


\author{B. Commer\c{c}on $^{1,2,}$}
\address{\'Ecole Normale Sup\'erieure de Lyon, Centre de recherche Astrophysique de Lyon (UMR 5574 CNRS), 
46 all\'ee d'Italie, 69364 Lyon Cedex 07, France}
 \address{Laboratoire AIM, CEA/DSM - CNRS - Universit\'e Paris Diderot,
DAPNIA/SAp, 91191 Gif sur Yvette, France}
\address{Laboratoire de radioastronomie millim\'etrique (UMR 8112 CNRS), \'Ecole Normale Sup\'erieure et Observatoire 
de Paris, 24 rue Lhomond, 75231 Paris Cedex 05, France}
\author{P. Hennebelle$^3$} 
\author{E. Audit$^2$}
\author{G. Chabrier$^1$}
\author{R. Teyssier$^2$}

\runningtitle{Numerical methods comparison for protostellar collapse calculations}

\setcounter{page}{1}

\index{Commer\c con B.}
\index{Hennebelle P.} 
\index{Audit E.}
\index{Chabrier G.}
\index{Teyssier R.}

\maketitle

\begin{abstract}
The development  of parallel supercomputers allows  today the detailed
study of the  collapse and the fragmentation of  prestellar cores with
increasingly accurate numerical simulations. Thanks to the advances in
sub-millimeter  observations,   a  wide  range   of  observed  initial
conditions enable  us to  study the different  modes of  low-mass star
formation.   The challenge  for the  simulations is  to  reproduce the
observational results. Two main numerical methods, namely AMR and SPH,
are  widely used  to simulate  the collapse  and the  fragmentation of
prestellar cores.  We compare here  thoroughly these two  methods with
numerical resolution requirements  deduced from previous studies.  Our
physical model is as simple as possible, and consists of an isothermal
sphere rotating around the  $z$-axis.  We first study the conservation
of angular momentum as a function of the resolution.  Then, we explore
a wide  range of simulation  parameters to study the  fragmentation of
prestellar cores.   There seems  to be a  convergence between  the two
methods, provided  resolution in each case  is sufficient.  Resolution
criteria adapted  to our  physical cases, in  terms of  resolution per
Jeans  mass,   for  an  accurate  description  of   the  formation  of
protostellar  cores   are  deduced  from  the   present  study.   This
convergence is encouraging for  future work in simulations of low-mass
star formation, providing the aforementioned criteria are fulfilled.
\end{abstract}

\section{Introduction}
Star formation is  known for being the place  of extreme variations in
length and density scales. Although  it is established that stars form
in dense cores, the non-linear evolution makes it difficult to perform
accurate  calculations of  the  collapse and  the  fragmentation of  a
prestellar core. The star formation  process is the outcome of complex
gas dynamics involving non-linear interactions of gravity, turbulence,
magnetic  field and  radiation.  Klein  et al.  (2007) point  out that
developing a  theory for  low-mass star formation  remains one  of the
most  elusive   and  important  goals   of  theoretical  astrophysics.
Numerical simulations  allow today  to study star  formation processes
and,  thanks   to  the  recent  development  of   super  computers,  a
significant  increase in  the description  of the  dynamical  range of
low-mass star formation has  been reached. The computational challenge
stems from  the fact  that star formation  occurs in clouds  over many
orders of magnitude in spatial  and density scales. The main issue for
the  simulations  is  to   follow  the  gravitational  collapse  while
resolving    precisely   the   Jeans    length,   which    scales   as
$\lambda_\mathrm{J} \propto \rho^{-1/2}$ for an isothermal gas.

Different  approaches  are  used   to  study  star  formation  through
numerical simulations and include  more and more detailed physics. One
key question resides  in the choice of the  most appropriate numerical
method to  be used  to study low-mass  star formation.   Nowadays, two
completely   different  approaches  can   handle  this   problem  with
sufficient accuracy: the Adaptive  Mesh Refinement method for Eulerian
grids (AMR) and the Smoothed Particle Hydrodynamics method (SPH) for a
Lagrangian approach.  We  use the AMR code RAMSES  (Teyssier 2002) and
the SPH code  DRAGON (Goodwin et al. 2004).   The debate about whether
one  method  is  most  appropriate  remains  open  and  no  systematic
comparison has  been done  in the domain  of low-mass  star formation.
Recently,  in a  comparative  study  between SPH  and  AMR, Agertz  et
al. (2006) have  shown that the SPH method is  of limited accuracy for
describing Kelvin-Helmotz  instabilities in  the presence of  a strong
density gradient. This effect in  SPH simulations is due mainly to the
cyclic, kick-drift  phenomenon, which strongly depends on  the form of
the  SPH force  calculation  coupled with  the  treatment of  neighbor
particles.   In the  low-mass  star formation  field,  Fromang et  al.
(2006) compares   quite  successfully   AMR   hydrodynamical  collapse
calculations  with SPH  ones.   Several studies  provide key  starting
points   for   the    numerical   investigation   of   low-mass   star
formation. Truelove et al. (1997)  give an empirical criterion for the
Jeans  length  resolution  in   AMR  calculations  to  avoid  spurious
numerical fragmentation while Bate \& Burkert (1997) provide a minimum
resolution  criterion  for   SPH  calculations  with  self-gravity  to
correctly  model  fragmentation.  In  the  present  study, we  compare
thoroughly the  two approaches in  the context of  low-mass prestellar
core  formation.   We focus  on  the  dependency  of angular  momentum
conservation  and  fragmentation  on  physical and  numerical  initial
conditions,  in order  to derive  resolution criteria  adapted  to our
study.

\section{Definitions of the test cases}

To  make comparison  between  codes easier,  we  adopt simple  initial
conditions,  similar to  those  chosen in  previous  studies (Bate  \&
Burkert  1997).  We  consider an  uniform sphere  of molecular  gas of
initial  radius $R_0$, rotating  around the  $z$-axis with  a constant
angular velocity  $\Omega_0$.  We set up  the cloud mass at  $M_0 = 1$
M$_{\odot}$ and  the temperature at 10  K.  For a  mixture of molecular
hydrogen,  helium  and  heavy  elements, with  mean  molecular  weight
$\mu=2.2$,  this   corresponds  to   an  isothermal  sound   speed  of
$C_\mathrm{0}   \sim  0.19   $  km.s$^{-1}$.    For  the   case  where
fragmentation  occurs,  we  use  a  $\mathrm{m}=2$  azimuthal  density
perturbation.  The  initial energy balance of our  model is determined
by two  dimensionless parameters  corresponding to the  ratio $\alpha$
between the  thermal energy  and the gravitational  energy and  to the
ratio $\beta$ of the rotational and the gravitational energy:
\begin{equation}
\alpha = \frac{5}{2}\frac{R_0kT}{GM_0\mu m_\mathrm{H}} \mbox{  ;  }\beta = \frac{1}{3}\frac{R_0^3\Omega_0^2}{GM_0}.
\end{equation}
Since we  use a constant initial  mass of 1 M$_{\odot}$  and a constant
temperature,  changing one  of  the two  parameters, namely  $\alpha$,
gives  the  sphere radius  $R_0$.   The  higher  $\alpha$, the  larger
$R_0$. The angular velocity is given by the parameter $\beta$.

In order to mimic the thermal  behaviour of a star-forming gas, we use
a  barotropic equation of  state. Masunaga  \& Inutsuka  (2000) showed
that  the  core  follows  closely  a  barotropic  equation  of  state,
providing   a   good   approximation   without   resolving   radiative
transfer. We use
\begin{equation}
\frac{P}{\rho} = C_\mathrm{s}^2 = C_0^2\left[1+\left(\frac{\rho}{\rho_c}\right)^{2/3}\right],
\label{baro}
\end{equation}
where  $C_0$ is  the isothermal  sound  speed at  10 K  and $\rho_c  =
10^{-13}$ g.cm$^{-3}$ is the critical density which corresponds to the
transition from an isothermal to  an adiabatic state (Larson 1969). At
low densities, $\rho \ll \rho_\mathrm{c}$, $C_\mathrm{s} \sim C_0=0.19
$ km.s$^{-1}$. The molecular gas is able to radiate freely by coupling
thermally to  the dust and therefore  remains isothermal at  10 K.  At
high densities  $\rho > \rho_\mathrm{c}$,  we assume that  the cooling
due to radiative transfer is  trapped by the dust opacity.  Therefore,
$P\propto \rho^{5/3}$ which corresponds to an adiabatic monoatomic gas
with adiabatic exponent $\gamma  = 5/3$.  Note that molecular hydrogen
behaves like  a monoatomic gas  until the temperature  reaches several
hundred  Kelvin,  since the  rotational  degrees  of  freedom are  not
excited  at  lower temperatures,  and  hence  $\gamma  = 5/3$  is  the
appropriate adiabatic exponent (e.g. Masunaga \& Inutsuka 2000).

\section{Numerical methods and refinement criteria}

\subsection{AMR code RAMSES}

In  this paper,  we use  the AMR  code RAMSES  (Teyssier  2002), which
integrates the  ``tree-based'' data structure  allowing recursive grid
refinements on a cell-by-cell  basis. RAMSES combines a tree-based AMR
grid and a  second order Godunov hydrodynamical scheme  coupled with a
gravity solver.  The Godunov  hydrodynamical solver is able to capture
discontinuities with  a high precision level. The  equations solved in
RAMSES   are  the   Euler  equations   in  their   conservative  form.
Furthermore, it has the possibility  to use variable timesteps at each
refinement   level.  Concerning  time   integration,  RAMSES   uses  a
second-order  midpoint  scheme,  where  positions and  velocities  are
updated by a predictor-corrector step.  Recently, an ideal MHD version
of RAMSES has been developed by Fromang et al. (2006). 

Our refinement criterion is based on the Jeans length resolution which
is necessary to treat  accurately gravitational collapse.  We impose a
minimum   number   of   points   $N_\mathrm{J}$   per   Jeans   length
$\lambda_\mathrm{J}$.   The cells  dimension must  be smaller  than a
constant fraction of  the local Jeans length.  The  dimension of cells
belonging     to     the      $\ell_i$     refinement     level     is
$L_\mathrm{box}/2^{\ell_i}$,  where $L_\mathrm{box}$  is  the physical
length of the simulation box.  The mesh is locally refined in order to
verify the local Jeans criterion:
\begin{equation}
\frac{L_\mathrm{box}}{2^{\ell_i}}< \frac{\lambda_\mathrm{J}}{N_\mathrm{J}}.
\end{equation}

Truelove et al. (1997) defined  a minimum resolution condition for the
validity of grid-based simulations aimed at modeling the collapse of a
molecular  cloud  core, namely  $N_\mathrm{J}  >  4$.  This  condition
ensures  that the  collapse is  of physical  rather than  of numerical
origin.

\subsection{SPH code DRAGON}

We use the  SPH code DRAGON (Goodwin et al.  2004).  The SPH formalism
relies  on an  interpolation method  which allows  any function  to be
expressed in  terms of  its values at  the location of  various points
called particles.   For numerical works, the  integral interpolant for
the  variable   $A(\textbf{r}_i)$  is  approximated   by  a  summation
interpolation over the particle's nearest neighbors:
\begin{equation}
  \label{eq:sum}
  A_s(\textbf{r}_i) = \sum_j m_j\frac{A(\textbf{r}_j)}{\rho
  (\textbf{r}_j)}W(|\textbf{r}_i-\textbf{r}_j|,h_{ij}),
\end{equation}
where   $A_j$   is   the   value   associated   with   particle   $j$,
$h_{ij}=(h_i+h_j)/2$  and $h_i$  is the  adaptive smoothing  length of
particle $i$,  defined such that  the particle kernel volume  (i.e the
resolution element) contains a  constant mass, i.e.  a constant number
of neighbors  $N_{N}$.  The  main advantage of  SPH is  its Lagrangian
conservation property in contrast with grid-based methods.  Resolution
elements are then concentrated in  high density regions in SPH methods
whereas the  AMR allows  high resolution of  all regions in  the flow.
The  standard   SPH  formalism  adopts   artificial  viscosity.   Some
alternative formalism such  as Godunov SPH has been  proposed in order
to avoid  the use of artificial  viscosity, but these  methods are not
yet mature.

In the SPH, the resolution in  mass is fixed and thus the Jeans length
resolution deteriorates with  increasing density during the isothermal
phase, contrary  to the AMR.  Bate  \& Burkert (1997)  showed that the
behaviour  of  a Jeans-mass  clump  of gas  with  radius  $\sim h$  is
dominated  by the  numerical implementation.   The  minimum resolvable
mass must then be larger  than the interpolation mass. Bate \& Burkert
(1997) take the smallest mass than can be resolved in SPH calculations
to be  equal to  the mass of  $\sim 2N_\mathrm{N}$ particles.   We can
determine  our  initial number  of  SPH  particles  according to  this
criterion.       The     Jeans      mass      is     $M_\mathrm{J}\sim
6\,G^{-3/2}\rho^{-1/2}C_\mathrm{s}^3$ and  the minimum resolvable mass
is $M_{res}=m  N_\mathrm{N}$, where $m$ is the  particle mass.  Hence,
we can define a Jeans  condition corresponding to the minimum value of
$C_\mathrm{s}^3\rho^{-1/2}$, given by the barotropic equation of state
(\ref{baro}), i.e. $2^{3/2}C_0^{3}\rho_c^{-1/2}$:
\begin{equation}
m < m_{max} \sim \frac{2^{3/2} 6\, C_0^{3}}{2N_\mathrm{N}
G^{3/2}\rho^{1/2}_c} \sim \frac{5.35 \times 10^{-3}}{N_\mathrm{N}} \mathrm{ M}_{\odot} .
\label{jeans_sph}
\end{equation}
Considering a spherical cloud of mass $M_0=1$ M$_{\odot}$, the initial
number of  particles $N_\mathrm{p}$ has to verify  $N_\mathrm{p} > M_0
/m_\mathrm{max} \sim 9300 $ if $N_\mathrm{N}=50$. This is the critical
number of particles used in  SPH calculations to study the collapse of
a dense core.  We have in that case  exactly $2N_\mathrm{N}$ (i.e. two
resolution elements) particles per critical Jeans mass.

\section{Results}

\subsection{Angular momentum conservation}
\begin{table}[htb]
\caption{Summary of the different  simulations (left table: SPH, right
table:  AMR)   performed  to  study   angular  momentum  conservation.
$N_\mathrm{i}$ for the  AMR calculations gives us the  number of cells
describing   the  initial   sphere.  The   quantity  $N_\mathrm{core}$
representing the number of cells/particles with density $\rho> 1\times
10^{-15}$ g.cm$^{-3}$. Time t$_0$ is the time when the density becomes higher than $\rho_\mathrm{c}=1\times 10^{-13}$ 
g.cm$^{-3}$.} \centering

\begin{tabular}{ccc}
\\
\begin{tabular}{ccccc}
\hline
\hline
$N_\mathrm{p}$    & $N_\mathrm{N}$ & $N_\mathrm{J} $ & $N_\mathrm{core}$& t$_0$ (kyr) \\
\hline
$5 \times 10^3$  & 50    & 1.86 & 225   & 115 \\
$1\times 10^4$   & 50    & 2.34 & 422   & 107 \\
$5\times 10^4$   & 50    & 4.   & 1 833 &  98 \\
$2\times 10^5$   & 50    & 6.35 & 7 055 &  95 \\
$5\times 10^5$   & 50    & 8.61 & 17 309&  93 \\
\hline
\end{tabular} 
&

&
\begin{tabular}{cccccc}
\hline
\hline
 $\ell_\mathrm{min}$   & $N_\mathrm{i}$ & $N_\mathrm{J}$  & $N_\mathrm{core}$ & Tot. cells & t$_0$ (kyr)\\
\hline 
5   &   2 145    &  6 &  3 928 & $\sim 9.1 \times 10^4$ & 150   \\
5   &   2 145    & 10 & 30 752 & $\sim 1.6 \times 10^5$ & 116    \\
6   &  17 160    &  4 &  4 016 & $\sim 3.1 \times 10^5$ & 116\\
6   &  17 160    & 10 & 28 800 & $\sim 3.7 \times 10^5$ & 109\\
7   & 137 260    & 10 & 29 944 & $\sim 2.2 \times 10^6$ &  96\\
\hline
\end{tabular}
\end{tabular}
\label{sum_homorot}
\end{table}

We start by comparing the global properties of the collapse in the two
codes in the simple case of  a uniform sphere rotating with a constant
angular velocity  around the $z$-axis, with no  perturbation.  We look
at  the  collapse time,  the  accretion  shock  (see Commer\c  con  et
al.  2007) and  finally the  angular momentum  conservation.   In this
section,  we focus on  this last  issue.  Considering  our axisymetric
model, without  azimuthal perturbation, we can  easily investigate the
effect of numerical resolution on local angular momentum conservation,
since local  angular momentum should be perfectly  conserved. The loss
of  local angular  momentum in  our model  is only  due  to unphysical
transport  inherent   to  the  numerical  methods  used   in  the  two
codes.Thanks to  its Lagrangian properties, the  SPH calculation gives
access  to the  angular  momentum that  each  particle has  initially,
i.e. the angular  momentum that particle should have  if the numerical
scheme  was  conserving it  exactly.  Having  access  to the  particle
initial angular momentum and to the same quantity at a given time, the
loss of angular momentum is easily calculated.

We carried out a set of  simulations within a wide range of resolution
parameters.   The initial sphere  has the  parameters $\alpha  = 0.65$
corresponding  to an  initial radius  $9.2  \times 10^{16}$  cm and  a
density   $\rho_0\sim   6.02   \times  10^{-19}$   g.cm$^{-3}$.    The
corresponding  freefall  time  is  t$_\mathrm{ff}  =  (3\pi  /  32  G
\rho_0)^{1/2} \sim 86 $ kyr.  We set $\beta=0.01$, corresponding to an
orbital   time   t$_\mathrm{rot}=2.8   \times  10^{3}$   kyr.    Table
\ref{sum_homorot}  summarizes the  different SPH  and  AMR calculation
runs for  this case. Calculations converged  with increasing numerical
resolution to a value  slightly greater than the theoretical freeffall
time because of the rotational support.

\begin{figure*}[htb]
  \centering
  \includegraphics[width=8cm,height=6cm]{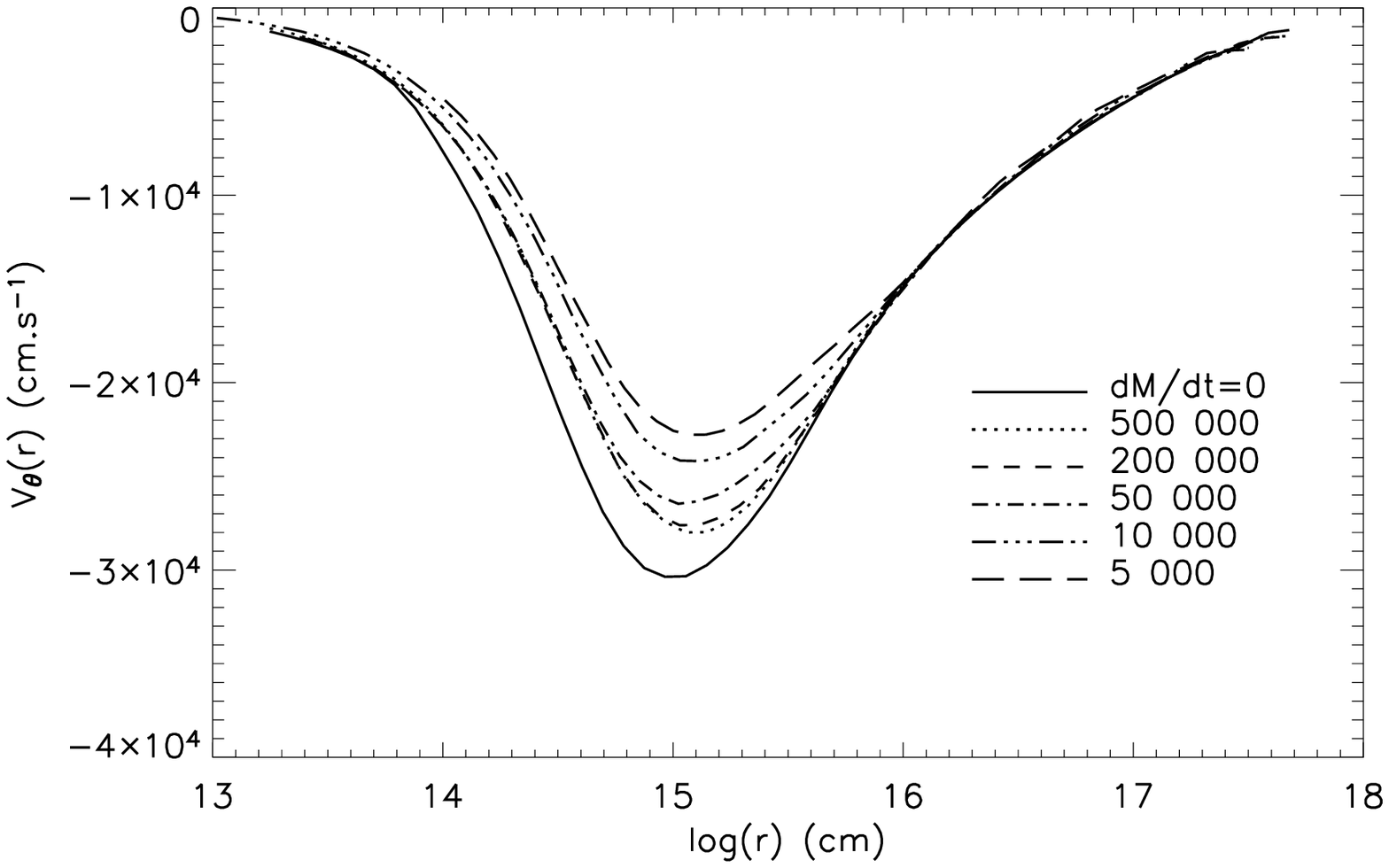}
  \includegraphics[width=8cm,height=6cm]{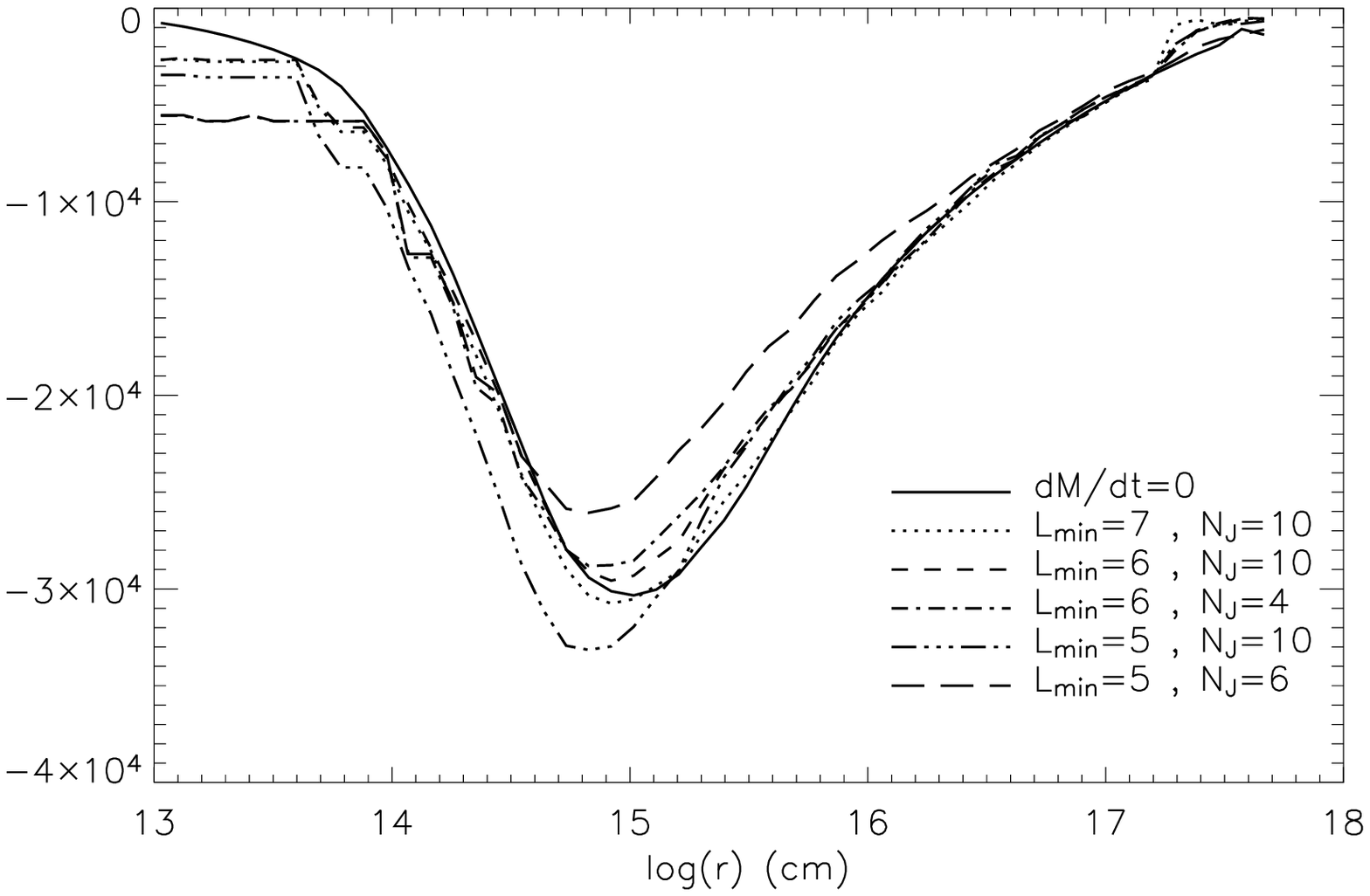}
  \caption{Azimuthal velocity at t$_0$ as  a function of the radius on
the equatorial  plane for SPH  (left) and AMR (right)  calculations at
corresponding t$_0$.   The left-hand plot  (Fig. \ref{homorot}a) shows
SPH    results   for   various    $N_\mathrm{p}$   and    a   constant
$N_\mathrm{N}=50$.    The  solid   line  represents   the  theoretical
azimuthal velocity interpolated at  t$_0$ and is denoted as $dM/dt=0$.
The  right-hand  plot  (Fig.  \ref{homorot}b) shows  AMR  results  for
$N_\mathrm{J}=10$ and $\ell_\mathrm{min}=5$, 6 and 7.  The theoretical
azimuthal velocity  is plotted also  for easy comparison with  the SPH
results.}
\label{homorot}
\end{figure*}

Figure \ref{homorot} (left)  displays the azimuthal velocity component
as  a  function of  the  radius  $r$ on  the  $xy$-plane  for the  SPH
simulations.  It  is obvious that  low resolution simulations  are not
able to  conserve properly the angular momentum.   With $5\times 10^3$
particles, we  obtain contra-rotating  particles at the  center.  This
pure numerical effect shows that a minimum resolution is required even
for a  simple collapse  model. It appears  that a minimum  of $5\times
10^4$  particles is  required to  get an  acceptable  angular momentum
conservation, within  less than $10 \%$,  for the case  of the present
study.  The improvement of angular momentum conservation with a larger
number  of  particles  eventually   saturates  for  large  numbers  of
particles.  We checked  that using a larger number  of neighbours does
not improve the conservation of local angular momentum.

Figure \ref{homorot}  (right) shows  results obtained with  the RAMSES
code. AMR  curves are  plotted and compared  with the  theoretical one
obtained   previously    for   the   SPH.     The   simulations   with
$\ell_\mathrm{min}=6$ and $\ell_\mathrm{min}=7$ give approximately the
same results.  In all  these AMR simulations,  the core  resolution is
better  than with the  SPH (see  parameter $N_\mathrm{core}$  in Table
\ref{sum_homorot}).      For      an     initial     resolution     of
$\ell_\mathrm{min}$=5,  the  AMR  scheme  does  produce  some  angular
momentum lag.   This can be due to  the fact that with  a poor initial
resolution, the interpolation of  the gravitational potential tends to
convert gravitational energy into rotational energy.

\subsection{Fragmentation}
This  subsection is devoted  to the  exploration of  various numerical
parameters in  a physical case known  to fragment. In  Commer\c con et
al. (2007), we study the effect of varying the initial grid resolution
$\ell_\mathrm{min}$  and the  number of  cells within  a  Jeans length
$N_\mathrm{J}$  for  AMR  calculations.    We  also  present  our  SPH
calculations with various  $N_\mathrm{N}$ and $N_\mathrm{p}$. We carry
out  this  study  for  three  different  cases  with  various  thermal
support.  Here,  we  focus   on  the  case  with  initial  parameters:
$\alpha=0.5$,   $\beta=0.04$,   giving  $\rho_0=1.35\times   10^{-18}$
g.cm$^{-3}$,   $R_0=7.07\times   10^{16}$   cm,   $\Omega_0=2.12\times
10^{-13}$  rad.s$^{-1}$  and   t$_\mathrm{ff}=57$  kyr.   The  initial
perturbation amplitude is $A=0.1$.

\begin{figure}[htb]
  \centering
    \includegraphics[width=8.cm,height=10.67cm]{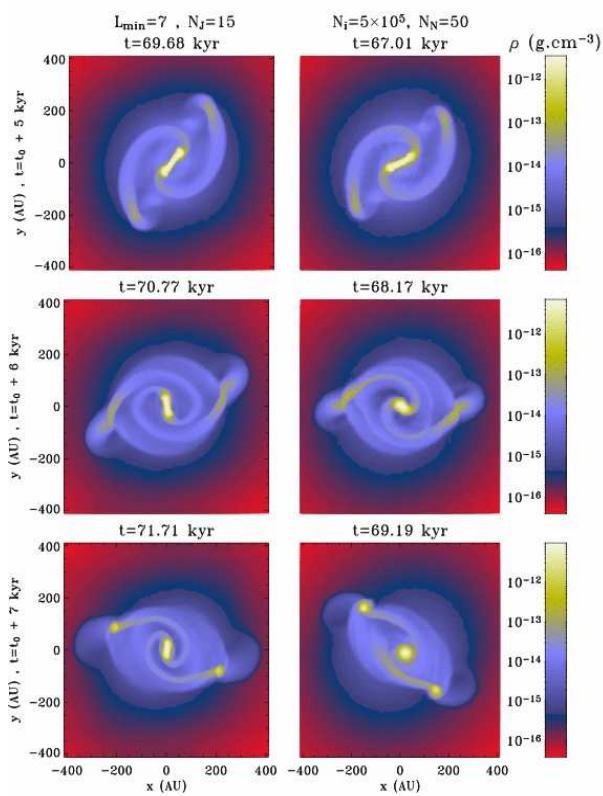}
  \caption{AMR and  SPH calculation density maps in  the $xy$-plane at
three different  times for the case  $\alpha=0.50$, $\beta=0.04$.  The
times correspond to  t$_0 + 5$ kyr, t$_0  + 6$ kyr and t$_0  + 7$ kyr,
from  to bottom, respectively.   The AMR  calculations plotted  on the
left   column   have   been   run   with   $\ell_\mathrm{min}=7$   and
$N_\mathrm{J}=15$.   The right column  shows the  results for  the SPH
calculations with $N_\mathrm{p}=5\times 10^5$ and $N_\mathrm{N}=50$.}
\label{comp_a050}
\end{figure}

Figure \ref{comp_a050} shows density  maps in the equatorial plane for
the results  of two  amongst the most  resolved calculations  at three
timesteps, namely,  from top to  bottom, t$_0$+5 kyr, t$_0$+6  kyr and
t$_0$+7  kyr. The  left column  shows maps  for AMR  calculations with
$\ell_\mathrm{min}=7$ and  $N_\mathrm{J}=15$ whereas the  right column
displays  SPH maps for  calculations with  $N_\mathrm{p}=5\times 10^5$
and $N_\mathrm{N}=50$. We show  here the two converged calculations of
each  method  that we  first  studied  independently. The  convergence
between the  two methods for  these physical and  numerical parameters
set is  striking for the  two first timesteps.  The  calculations give
the same  fragmentation time and pattern, although  satellites and the
central object are bigger with  the SPH. Symmetry is eventually broken
with the SPH calculations, whereas  the AMR ones preserve the symmetry
longer.  

\section{Conclusion}
 
We show  that we  reach good convergence  between AMR and  SPH methods
provided one  uses sufficient numerical  resources.  First, we  take a
simple model to study local angular momentum conservation. The initial
study shows  that angular  momentum is better  conserved with  the AMR
approach.   As shown  in  Fig.  \ref{homorot}a,  a  smaller number  of
particles  in SPH calculations  leads to  poor local  angular momentum
conservation. In AMR calculations, a poor initial computational domain
resolution  (i.e.   $\ell_\mathrm{min}   <  6$)  leads  to  unphysical
transfer  of  gravitational  energy  to rotational  energy  (see  Fig.
\ref{homorot}b). A  significant loss  of angular momentum  will affect
fragmentation since less  rotational support can balance gravitational
collapse.  The  smallest parameter set for  SPH calculations, required
to  go  through gravitational  collapse  without  significant loss  of
angular momentum, corresponds to a  number of $\sim 530$ particles per
Jeans  mass  at  the   critical  density  $\rho_\mathrm{c}$,  i.e.   5
particles per  Jeans {\it length}.  The  equivalent minimum resolution
criterion  for  AMR  calculations   is  $\ell_\mathrm{min}  >  6$  and
$N_\mathrm{J}=4$.

For the fragmentation case, the two approaches show good agreement for
the general pictures. Details  are better resolved in AMR calculations
thanks  to the  refinement method  based  on the  local Jeans  length,
whereas the  resolution deteriorates with increasing  density with the
SPH, because of the  fixed mass resolution.  Numerical calculations of
protostellar collapse should thus be conducted with great care, with a
detailed examination of numerical resolution.  The resolution criteria
to  reach convergence for  the cases  we explore  in details  is about
$\sim 5000$  particle per Jeans  mass in SPH and  $N_\mathrm{J}=15$ in
AMR, and could become prohibitive  in some cases, particularly for low
thermal  support. In  such  a case,  it  is more  difficult  to get  a
convergence.   The  horizon of  predictability  is  very short,  since
fragmentation occurs very quickly.   The latter parameter set seems to
be a lower resolution limit  for dense core collapse and fragmentation
with SPH calculations. Using a  lower number of particles will lead to
spurious early  fragmentation due  to numerical effects.   The present
work can  be used to assess  the validity of numerical  tools to study
star formation.

\begin{acknowledgements}
Calculations have  been performed  thanks at the  PSMN (ENS  Lyon) and
CCRT  (CEA)  supercomputating facilities,  as  well  as  on the  CEMAG
computing facility  supported by the  French ministry of  research and
education through a Chaire d'Excellence awarded to Steven Balbus.
\end{acknowledgements}

{}

\end{document}